\documentclass[conference]{IEEEtran}

\usepackage{graphicx}
\usepackage{listings}
\usepackage{enumitem}
\usepackage{subcaption}
\usepackage{todonotes}
\usepackage{verbatim}

\graphicspath{{images/}}
\DeclareGraphicsExtensions{.pdf,.jpeg,.png,.jpg,.eps}

\begin{document}
%
\title{Version 0.1 of the BigDAWG Polystore System}



%
\author{\IEEEauthorblockN{Vijay
    Gadepally\IEEEauthorrefmark{1}\IEEEauthorrefmark{2} Kyle O'Brien\IEEEauthorrefmark{1}  Adam 
    Dziedzic\IEEEauthorrefmark{4}  Aaron Elmore\IEEEauthorrefmark{4}
     Jeremy Kepner\IEEEauthorrefmark{1}\IEEEauthorrefmark{2} \\
    Samuel Madden\IEEEauthorrefmark{2} Tim
    Mattson\IEEEauthorrefmark{5} Jennie Rogers\IEEEauthorrefmark{3}
    Zuohao She\IEEEauthorrefmark{3} Michael
    Stonebraker\IEEEauthorrefmark{2}} \\
\IEEEauthorblockA{\IEEEauthorrefmark{1}MIT Lincoln Laboratory
    \IEEEauthorrefmark{2}MIT CSAIL   \IEEEauthorrefmark{3}Northwestern
  University   \IEEEauthorrefmark{4}University of Chicago
    \IEEEauthorrefmark{5}Intel Corporation}}

\maketitle

\let\thefootnote\relax\footnotetext{The corresponding author, Vijay Gadepally, can be reached at
 \\ vijayg [at] ll.mit.edu.}

\begin{abstract}


A polystore system is a database
management system (DBMS) composed of integrated heterogeneous
database engines and multiple programming languages. 
By matching data to the storage engine best suited to its needs,
complex analytics run faster and flexible storage choices helps
improve data organization.  BigDAWG (Big Data Working Group) is our reference 
implementation of a polystore system. In this paper, we
describe the current BigDAWG software release which supports PostgreSQL,
Accumulo and SciDB. We describe the overall architecture, API and
initial results of applying BigDAWG to the MIMIC II medical dataset.




\end{abstract}


\IEEEpeerreviewmaketitle

\section{Introduction}

%
%
%
%

Data comes in all shapes and sizes and it is unlikely that any single
database management system will be able to efficiently query and analyze 
diverse data such as text, images, graphs,
video, etc~\cite{stonebraker2005one}. Thus, the need for
database systems that leverage heterogenous data stores
such as relational systems, key-value stores, graph databases,
in-memory databases, array databases, etc. 

Polystore systems are of great interest across the research community~\cite{begoli2016towards}. 
They integrate diverse database
engines and multiple programming languages presenting them as a 
single system. The BigDAWG
system~\cite{gadepally2016bigdawg} is our implementation of a polystore database. 
BigDAWG's architecture consists of four distinct layers: database and
storage engines; islands; middleware and API; and applications. Our
previous results described the development of core BigDAWG features,
and its application to medical~\cite{elmore2015demonstration} and
scientific datasets~\cite{mattsondemonstrating}.  In this paper, we
describe the open source release of BigDAWG (available at
http://bigdawg.mit.edu under the BSD-3 license) and an analysis of the performance of 
BigDAWG queries applied to
the MIMIC II medical dataset~\cite{saeed2011multiparameter}.

The remainder of the article is organized as
follows: Section~\ref{polystore} expands on the concept of a polystore
databases and the execution of polystore
queries; Section~\ref{bigdawg} discusses the overall architecture of
the BigDAWG system; Section~\ref{release} discusses the specifics of
our recent software release; Section~\ref{internals:middleware-components} describes the current BigDAWG
architecture and its application to the MIMIC II
dataset, and Section~\ref{results} describes performance results on an
initial BigDAWG implementation. Finally, we conclude and discuss future
work in Section~\ref{conc}.

%

\section{Polystore Systems}
\label{polystore}

The ``one size does not fit all''~\cite{stonebraker2005one} slogan is now famous in the database community.  If data storage engines
match the data, performance of data intensive applications
is greatly enhanced. In our previous work, we show that such
benefits can often lead to orders of
magnitude performance advantages~\cite{gadepally2016bigdawg}. Beyond performance, organizations may already have data spread
across a number of storage engines and the data must reside in those
original systems for policy or performance reasons. Writing connectors to move queries
and data across $N$
different systems could require up to $O(N^{2})$ connectors, a fact that
complicates adoption of polystore techniques.

A polystore system is a database management system (DBMS) that is built on top of multiple, heterogeneous, integrated
storage engines.  Each of these terms is important to distinguish a
Polystore from conventional federated DBMS.

By our definition, a polystore must consist of \textbf{multiple} data stores. However, polystores should not to be confused with a distributed DBMS
which consists of replicated or partitioned instances of a storage engine sitting behind a single query engine.  The key to a polystore
is that the multiple storage engines are distinct and accessed
separately through their own query engine.

Therefore, storage engines must be \textbf{heterogeneous} in a polystore system. If
they were the same, it would violate the whole point of polystore systems;
i.e., the mapping of data onto distinct storage engines well suited to the
features of components of a complex data set.

Finally, the storage engines must be \textbf{integrated}. In a federated DBMS, the individual storage engines are independent.
In most cases, they are not managed by a single administration team.  In a polystore system,
the storage engines are managed together as an integrated set.  The challenge in designing a polystore system is to balance two often conflicting forces.
\begin{itemize}
\item {} 
\emph{Location Independence}: A query is written and the system figures out which storage engine it targets.

\item {} 
\emph{Semantic Completeness}: A query can exploit the full set of features provided by a storage engine.

\end{itemize}

BigDAWG is our reference implementation of a polystore system. It is by no means, however, the only 
such system as a number of groups are also exploring different approaches to 
polystore DBMS systems~\cite{spyropoulos2016digree,maccioni2016quepa,kharlamov2016semantic,dasgupta2016analytics}. The
remainder of this article concentrates on the open source release of the
BigDAWG system and a range implementation details.

\section{BigDAWG}
\label{bigdawg}

\begin{figure}[htbp]
\centering
\includegraphics[width=2.5in]{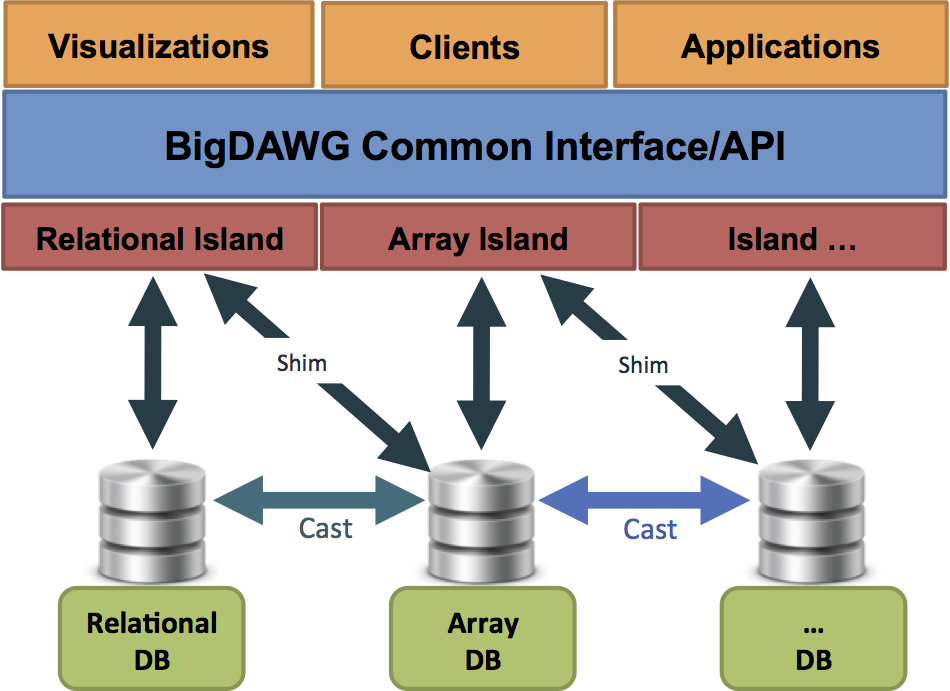}
\caption{BigDAWG Architecture. \label{bigdawgArchitecture}}
\end{figure}

Figure~\ref{bigdawgArchitecture} describes the overall BigDAWG architecture.  At the bottom, we have a collection
of disparate storage engines.  These engines are organized into a number of \emph{islands}.
An island is composed of a data model, a set of operations and a set
of candidate storage engines.  An
island provides location independence among its associated storage
engines.

A \emph{shim} connects an island to one or more storage engines.  The shim
translates queries expressed in terms of the
island's operations into the native query language of
a particular storage engine. For example, the \texttt{relational}
island may convert a BigDAWG query into an \textit{SQL} query that
can be understood by PostgreSQL, MySQL or other relational database.

A key goal of a polystore system is for processing to occur on the
storage engine best suited to the features of the data.  We expect
in typical workloads that queries will produce results best suited to
particular storage engines.  Hence, BigDAWG needs a capability to move
data directly between storage engines.  We do this with software
components we call \emph{casts}.

The BigDAWG common interface is middleware that supports a common
application programming interface to a collection of storage engines
via a set of islands. The middleware consists of a number of components:

\begin{itemize}
\item Optimizer: parses the input query and creates a set of viable query plan trees with possible engines for each subquery.
\item Monitor: uses performance data from prior queries to determine the query plan tree with the best engine for each subquery.
\item Executor: figures out how to best join the collections of objects and then executes the query.
\item Migrator: moves data from engine to engine when the plan calls for such data motion.
\end{itemize}

\begin{figure}[tbp]
\centering
\includegraphics[width=3.4in]{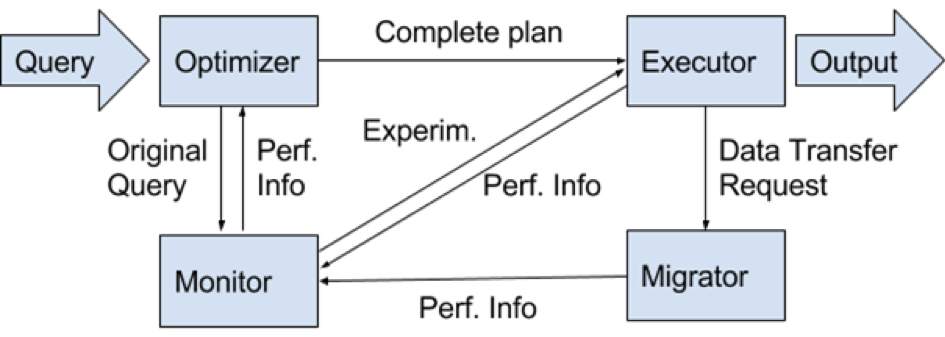}
\caption{Internal Components of the BigDAWG Middleware.}
\label{bigdawgmiddleware}
\end{figure}

These components are shown graphically in
Figure~\ref{bigdawgmiddleware}. Given an incoming query, the
planner parses the query into collections of objects and creates a set
of possible query plan trees over collections of engines
and objects. The planner  sends these trees to
the monitor which uses existing performance information to determine a
tree with the best engine for each collection of objects (based on
previous experience of a similar query). The tree is then passed to
the executor which determines the best method to combine the
collections of objects and executes the query. The executor can use
the migrator to move objects between engines and islands, if required,
by the query plan. 

\section{Current BigDAWG Release}
\label{release}

\begin{figure}[htbp]
\centering
\includegraphics[width=3.4in]{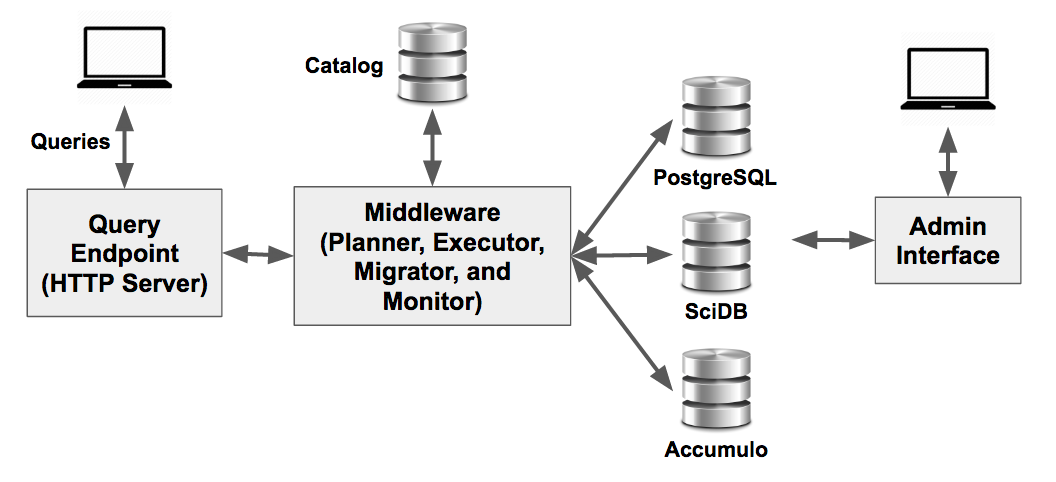}
\caption{BigDAWG software release system overview.}
\label{system_overview}\end{figure}

In March 2017, we released BigDAWG version 0.1 licensed under
the terms of the BSD license (http://bigdawg.mit.edu). This initial BigDAWG release supports three
open-source database engines: PostgreSQL (SQL), Apache Accumulo
(NoSQL), and SciDB (NewSQL) along with support for relational, array
and text islands. The middleware is packaged in Docker containers. Figure~\ref{system_overview} describes the software
components of the BigDAWG software release.

Users primarily interact with the Query Endpoint, which accepts
queries, routes them to the Middleware, and responds with results. The
Catalog is a PostgreSQL engine containing metadata about the other
engines, datasets, islands and connectors managed by the
Middleware.

The middleware is modular and allows additional islands or database engines
to be supported. Soon, we
will also support MySQL, the columnar store
Vertica~\cite{lamb2012vertica}, and the streaming engine
S-Store~\cite{cetintemel2014s}. 

In addition to the core software, we developed
polystore solutions for medical~\cite{elmore2015demonstration} and scientific~\cite{mattsondemonstrating}
data. The current BigDAWG release
includes scripts to download and load the MIMIC II medical
dataset~\cite{saeed2011multiparameter}. Using the default settings,
patient history data is inserted into PostgreSQL, physiologic waveform
data is inserted to SciDB, and free-form text data is inserted into
Accumulo. We also include a number of
example BigDAWG queries and an administrative interface to start, stop and
view the status of a BigDAWG setup.

\section{BigDAWG Components}
\label{internals:middleware-components}

As shown in Figure~\ref{bigdawgArchitecture}, BigDAWG is constructed from a number
of components. In this paper we describe the
middleware responsible for receiving
queries, translating them into statements that execute on
actual data stores, and maintaining information about the system configuration. 
Other BigDAWG components are described in~\cite{gadepally2016bigdawg}.

The BigDAWG middleware consists of four modules: query planning,  performance monitoring, data
migration and query execution. Information about the hardware
configuration is maintained in the catalog. We
describe these BigDAWG modules below. 

\subsection{Catalog}
\label{internals:catalog}
The Catalog stores metadata about the system.
The Planner, Migrator, and Executor all rely on the Catalog for
``awareness'' of the BigDAWG components, such as the hostname and IP
address of each engine, Engine to Island assignments, and the data
objects stored on each engine.

The Catalog is maintained by a PostgreSQL instance with a database 
called the \texttt{bigdawg\_catalog}. 
%
This database is made up of a number of tables which define items currently managed by the BigDAWG middleware:
\begin{itemize}
\item \texttt{engines}: Engine names and connection information.
\item \texttt{databases}: Databases, their engine membership, and connection authentication information.
\item \texttt{objects}: Data objects (\emph{i.e.}, tables), field-names, and object-to-database membership.
\item \texttt{shims}: Engines integrated with each island.
\item \texttt{casts}: the available casts between engines.
\end{itemize}

\begin{figure*}[h]
\centering
\begin{subfigure}{0.5\textwidth}\centering
\includegraphics[width=\linewidth]{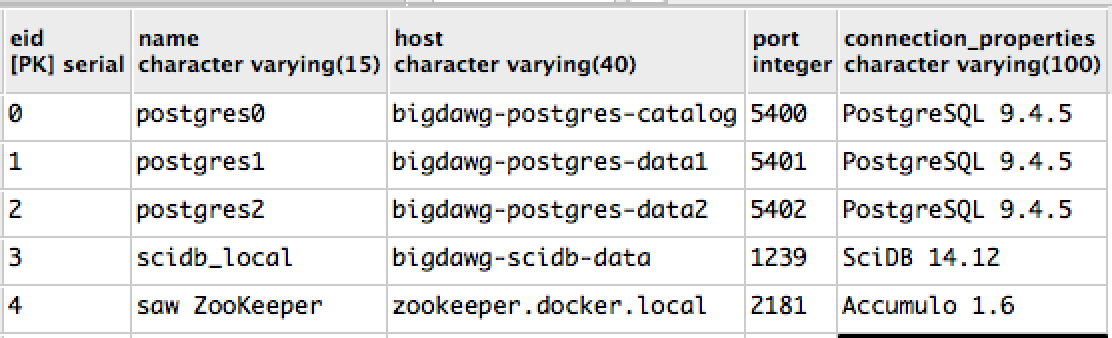}
\caption{Example Engines Table}\end{subfigure}\hfill 
\begin{subfigure}{0.45\textwidth}\centering
\includegraphics[width=\linewidth]{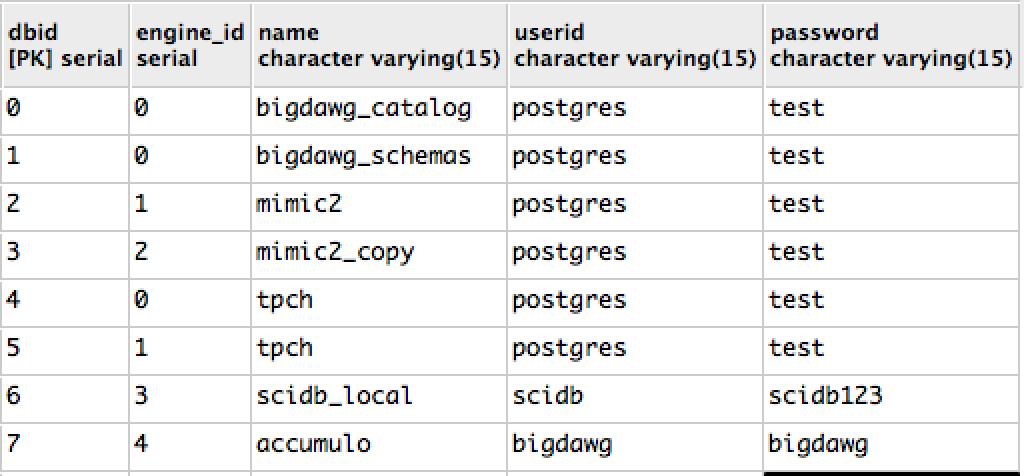}
\caption{Example Databases Table}\end{subfigure} \bigskip \hfill
\begin{subfigure}{0.6\textwidth}\centering
\includegraphics[width=\linewidth]{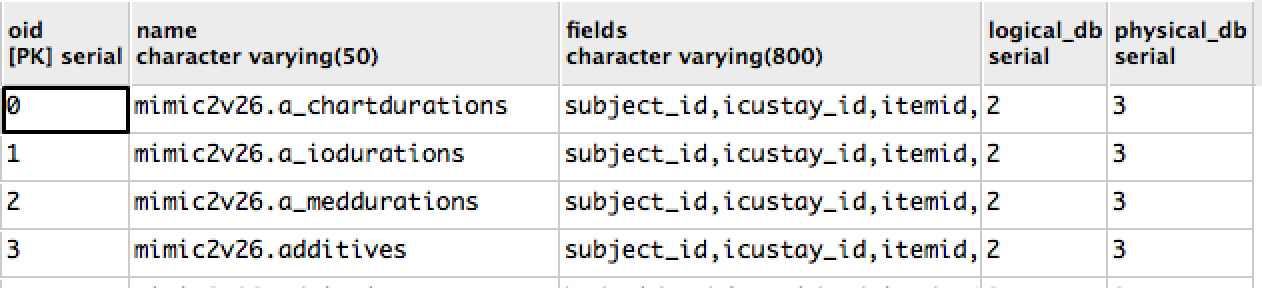}
\caption{Example Objects Table}\end{subfigure}\hfill
\begin{subfigure}{0.35\textwidth}\centering
\includegraphics[width=\linewidth]{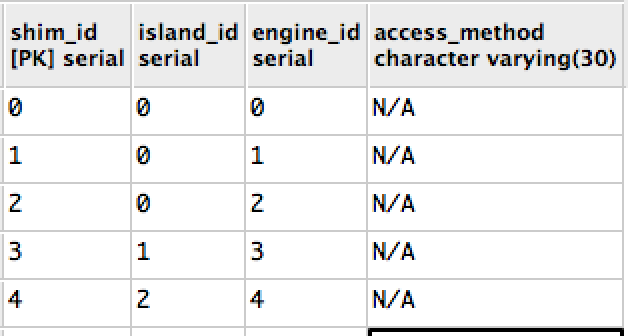}
\caption{Example Shims Table}\end{subfigure}
\caption{BigDAWG catalog tables contain information about what
  engines, databases, tables and data objects are being managed by the
middleware.}
\label{catalog_table}
\end{figure*}

Examples of contents of these tables are given in
Figure~\ref{catalog_table}. In order for BigDAWG to ``see'' new engines, databases or
objects a user must add entries to relevant tables in the
Catalog. BigDAWG includes scripts and an
administrative interface to simplify modifications to the Catelog.

\subsection{Planner}
\label{internals:planner}
The Planner~\cite{she2016bigdawg} coordinates all query execution.
It has a single static function that initiates query processing for a
given query and handles the result output. A relevant fragment of code
from the Query Planning module is shown below:

\begin{lstlisting}[language=Java,frame=none]
package istc.bigdawg.planner;

public class Planner {
    public static Response processQuery(
        String userinput, boolean isTrainingMode
    ) throws Exception
}
\end{lstlisting}

The String \texttt{userinput} is the BigDAWG query. 
The \textit{processQuery()} function first checks if the query 
is intended to interact with the Catalog.
If so, the query is routed to a special processing module to 
parse and process Catalog-related queries.
Otherwise, \texttt{processQuery()}  parses and processes the query string. 
When the boolean of \texttt{isTrainingMode} is \texttt{true}, 
the Planner performs query optimization by enumerating 
all possible orderings of execution steps
that produce an identical result. 
Then, the Planner sends the enumeration to the Monitor to gather query execution metrics.
The Planner then picks the fastest plan to run returning the result to the Query Endpoint.
When \texttt{isTrainingMode} is \texttt{false}, the Planner
consults the Monitor to retrieve the best query plan based on past
execution metrics. 

Data retrieval queries are passed as inputs to the constructor of a \texttt{CrossIslandQueryPlan} object.
A \texttt{CrossIslandQueryPlan} object holds a nested structure that represents a plan for inter-island query execution.
An inter-island query execution is specified by \texttt{CrossIslandPlanNode} objects organized in tree structures: the nodes either carry information
for an intra-island query or an inter-island migration.

Following the creation of the \texttt{CrossIslandQueryPlan}, the Planner traverses the tree structure of \texttt{CrossIslandPlanNode} objects and executes the
intra-island queries, invokes migrations, and then produces the final result.

\subsection{Migrator}
\label{internals:migrator}
The BigDAWG data migration module~\cite{dziedzic2016data} exposes a single interface to other
modules. Clients provide the connection information for source and destination databases as well as
a name of the object (e.g. table, array) to be extracted from the source database, and a name of the object
(e.g. table, array) to which the data should be loaded.

\begin{lstlisting}[language=Java,frame=none]
package istc.bigdawg.migration;

public class Migrator {
   public static MigrationResult migrate(
     ConnectionInfo connectionFrom, String objectFrom,
     ConnectionInfo connectionTo, String objectTo,
     MigrationParams migrationParams)
        throws MigrationException;
   }
 }
\end{lstlisting}

Internally, the Migrator identifies the type of the databases by examining the connection information.
The \texttt{ConnectionInfo} object is merely an
\emph{interface} so we check the actual type of the object.
The connection object represents a specific database (e.g.
PostgreSQL, SciDB, Accumulo or S-Store).
Currently, we support migration between instances of PostgreSQL,
SciDB, and Accumulo. There is an efficient binary
data migration between PostgreSQL and SciDB.
We also support binary migration to Vertica, but Vertica does not expose binary export. 
Future work will build a distributed migrator,
tighter integration with S-Store, and a more
efficient connection with Accumulo.

\subsection{Executor}
\label{internals:executor}
The Executor~\cite{gupta2016cross} executes intra-island queries through static functions.
The static functions create instances of \texttt{PlanExecutor} objects
that execute individual intra-island queries. A fragment of the
executor definition is shown below:

\begin{lstlisting}[language=Java,frame=none]
package istc.bigdawg.executor;

public class Executor {
    public static QueryResult executePlan(
        QueryExecutionPlan plan,
        Signature sig,
        int index
    ) throws ExecutorEngine.LocalQueryExecutionException, MigrationException;

    public static QueryResult executePlan(
        QueryExecutionPlan plan
    ) throws ExecutorEngine.LocalQueryExecutionException, MigrationException;

    public static CompletableFuture<Optional<QueryResult>> executePlanAsync(
        QueryExecutionPlan plan,
        Optional<Pair<Signature, Integer>> reportValues
    );
}
\end{lstlisting}

The \texttt{PlanExecutor} objects are created from
\texttt{QueryExecutionPlan} objects that represent execution
plans of an intra-island query. A \texttt{QueryExecutionPlan}, QEP,
holds details of sub-queries required for their execution and
a graph of dependency information among the
sub-queries. The \texttt{PlanExecutor} takes information from a
\texttt{QueryExecutionPlan} object and issues the sub-queries to their
corresponding databases and calls the appropriate \texttt{Migrator} classes to migrate intermediate results.

\subsection{Monitor}
\label{internals:monitor}
The monitor~\cite{chen2016bigdawg} manages queries and records
query performance of the BigDAWG system. A fragment of
the monitor definitions is shown below:

\begin{lstlisting}[language=Java,frame=none]
class Monitor {
  public static boolean addBenchmarks(Signature signature, boolean lean);
  public static List<Long> getBenchmarkPerformance(Signature signature);
  public static Signature getClosestSignature(Signature signature);
 }
\end{lstlisting}

The \texttt{signature} parameter identifies a query.  The
\texttt{addBenchmarks} method adds a new benchmark. If the
\texttt{learn} parameter is \texttt{false}, the benchmark is
immediately run over all possible query execution plans
(henceforth referred to as QEP). The \texttt{getBenchmarkPerformance} method returns a list of
execution times for a particular benchmark, ordered in the same order that
the benchmark's QEPs are received.  The best way to use the module is to add all of the relevant benchmarks 
using the \textit{addBenchmarks} method and then
retrieve information through \texttt{getBenchmarkPerformance}.

One of the more useful features is contained in the \texttt{getClosestSignature} method, which tries to find the closest matching benchmark for
the provided signature. A user can add benchmarks that cover the majority of query use cases and then
use the \texttt{getClosestSignature} method to find a matching benchmark and compare the QEP times to those of the current signature.  
If no matching signatures are found, the current signature is added as a new benchmark.

There are many opportunities to enhance this feature to improve the
matching, possibly by utilizing machine learning techniques. The
public methods in the \texttt{Monitor} class are the only API
endpoints that should be used. In contrast, the
\texttt{MonitoringTask} class updates the benchmark timings
periodically and should be run in the background through a daemon.

\section{BigDAWG API}
\label{api}

BigDAWG queries are written with the BigDAWG Query language (BQL)
which uses a functional syntax. At the highest level, a BigDAWG query 
follows the following format:

\begin{lstlisting}
bdrel( ... )
\end{lstlisting}

A function token (`bdrel' in this case) indicates how the syntax
within the parenthesis is interpreted.  For example,
the `bdrel' function token indicates that this is a
query for the relational island and any code between the parenthesis
will be interpreted as SQL code.

Five function tokens are defined in BigDAWG.  Three function tokens
are used to issue queries/sub-queries within individual islands:
\begin{itemize}
\item \texttt{bdrel} -- the query targets the relational island and uses PostgreSQL.
\item \texttt{bdarray} -- the query targets the array island and uses SciDB's AFL query language.
\item \texttt{bdtext} -- the query targets the text island and uses either SQL or D4M.
\end{itemize}

One function token deals with data migration between islands:
\begin{itemize}
\item \texttt{bdcast} -- the query is a cast operation for inter-island data migration.
\end{itemize}

The remaining function token handles metadata for the polystore system:
\begin{itemize}
\item \texttt{bdcatalog} -- the query targets the BigDAWG catalog using SQL.
\end{itemize}

The island and migration tokens are often nested within each other,
while the \texttt{bdcatalog} token is only usable on its own.  
Sub-queries using the \texttt{bdcast} function token are
always nested between a pair of island queries.

\subsection{Syntax Definitions}
\label{query-language:bigdawg-query}

BigDAWG Query Syntax:

\begin{lstlisting}[language=bash,caption={BigDAWG Query Syntax}]
BIGDAWG_SYNTAX ::=
  BIGDAWG_RETRIEVAL_SYNTAX | CATALOG_QUERY

BIGDAWG_RETRIEVAL_SYNTAX ::=
  RELATIONAL_ISLAND_QUERY | ARRAY_ISLAND_QUERY | TEXT_ISLAND_QUERY
\end{lstlisting}

\paragraph{Catalog Manipulation}
\label{query-language:catalog-manipulation}\label{query-language:id1}
Catalog manipulation queries are used to directly view the content
of the catalog.

\begin{lstlisting}[language=bash,caption={Catalog Manipulation Syntax}]
CATALOG_QUERY ::=
  { bdcatalog( catalog_table_name { [ column_name ] [, ...] }) }
  | { bdcatalog( query_applied_to_catalog_database ) }
\end{lstlisting}

\paragraph{Relational Island}
\label{query-language:relational-island}
The Relational Island follows the relational data model with 
data organized into tables. The rows of a table are
termed as \emph{tuples} and columns simply as \emph{columns}.

The Relational Island currently supports a subset of
SQL used by PostgreSQL. It allows for single-layered
\textit{SELECT} query with filter, join, aggregation, sort and limit
operations. Column projection, simple arithmetic SQL expression and 
basic aggregate functions (i.e. count, sum, avg, min and max) 
are supported. 

\begin{lstlisting}[language=bash,caption={Relational Island Syntax}]
RELATIONAL_ISLAND_QUERY ::=
  bdrel( RELATIONAL_SYNTAX )

RELATIONAL_SYNTAX ::=
  SELECT [ DISTINCT ]
  { * | { SQL_EXPRESSION [ [ AS ] output_name ] [, ...] } }
  FROM FROM_ITEM [, ...]
  [ WHERE SQL_CONDITION ]
  [ GROUP BY column_name [, ...] ]
  [ ORDER BY SQL_EXPRESSION [ ASC | DESC ]
  [ LIMIT integer ]

FROM_ITEM ::=
  { table_name | BIGDAWG_CAST } [ [ AS ] alias ]
\end{lstlisting}

The following is an example of a relational island query; it uses the 
relational island (\texttt{bdrel}) to select 4 entries from the 
table \texttt{mimic2v26.d\_patients}.

\begin{lstlisting}[language=bash,frame=none]
bdrel(select * from mimic2v26.d_patients limit 4)
\end{lstlisting}

\paragraph{Array Island}
\label{query-language:array-island}
The Array Island follows an array data model, where data is organized
into arrays. Arrays are multi-dimensional grids, where each cell in
the grid contains a number of fields. Each dimension of an array is
referred to as a \emph{dimension} and each field in a cell is termed an
\emph{attribute}. Dimensions assume unique values whereas attributes are
allowed duplicates. A combination of dimension values across all
dimensions in an array uniquely identify an individual cell of attributes.

The Array Island currently supports a subset of SciDB's Array
Functional Language (AFL). It allows for project, aggregation,
cross\_join, filter and schema reform (redimension). The Array Island also allows
attribute sorting; however, at the moment, only sort in ascending
order is supported.

\begin{lstlisting}[language=bash,caption={Array Island Syntax}]
ARRAY_ISLAND_QUERY ::=
  bdarray( ARRAY_SYNTAX )

ARRAY_SYNTAX ::=
  { scan( ARRAY_ISLAND_DATA_SET ) }
  | { project( ARRAY_ISLAND_DATA_SET [, attribute ] [...]) }
  | { filter( ARRAY_ISLAND_DATA_SET, SCIDB_EXPRESSION ) }
  | { aggregate( ARRAY_ISLAND_DATA_SET, SCIDB_AGGREGATE_CALL [, ...] [, dimension] [...] ) }
  | { apply( ARRAY_ISLAND_DATA_SET {, new_attribute, SCIDB_NON_AGGREGATE_EXPRESSION} [...] ) }
  | { cross_join( ARRAY_ISLAND_DATA_SET [ as left-alias], ARRAY_ISLAND_DATA_SET [ as right-alias ] [, [left-alias.]left_dim1, [right-alias.]right_dim1] [...] ) }
  | { redimension( ARRAY_ISLAND_DATA_SET, { array_name | SCIDB_SCHEMA_DEFINITION } ) }
  | { sort( ARRAY_ISLAND_DATA_SET [, attribute] [...] } ) }

ARRAY_ISLAND_DATA_SET ::=
  array_name | ARRAY_ISLAND_SYNTAX | BIGDAWG_CAST
\end{lstlisting}

An example array island query is shown below; it filters all entries in the array \texttt{myarray} with dimension \texttt{dim1} greater than $150$. 

\begin{lstlisting}[language=bash,frame=none]
bdarray(filter(myarray,dim1>150))
\end{lstlisting}

\paragraph{Text Island}
\label{query-language:text-island}
The Text Island logically organizes data in tables, and retrieves
data in a key-value fashion. This is modeled after the data model of
the Accumulo engine. When queried for a certain table, it returns
a list of key-value pairs. The key contains row label, column family label,
column qualifier label, and a time stamp. The value is a string.

The Text Island query syntax adopts a JSON format using single-quote
for labels and entries. The user can issue full table scan or
range retrieval queries.

\begin{lstlisting}[language=bash,caption={Text Island Syntax}]
TEXT_ISLAND_QUERY ::=
  bdtext( TEXT_ISLAND_SYNTAX )

TEXT_ISLAND_SYNTAX ::=
  { 'op' : 'TEXT_OPERATOR', 'table' : '(table_name | BIGDAWG_CAST)' [, 'range' : { TEXT_ISLAND_RANGE }] }
\end{lstlisting}

An example text island query is shown below; this query illustrates the text island (\texttt{bdtext}) to scan all entries in the Accumulo table \texttt{mimic\_logs} with row keys between \texttt{r\_0001} and \texttt{r\_00015}, matching any column family and column qualifier.

\begin{lstlisting}[language=bash,frame=none]
bdtext({ 'op' : 'scan', 'table' : 'mimic_logs', 'range' : { 'start' : ['r_0001','',''], 'end' : ['r_0015','','']} })
\end{lstlisting}

\paragraph{Inter-Island Cast}
\label{query-language:inter-island-cast}
Inter-island casts move data between different
islands. The differences between two data models can give rise to ambiguities
when migrating data between them. When issuing a Cast that invokes
an Inter-Island migration, the user avoids such ambiguities by
providing the schema used in the destination island.

\begin{lstlisting}[language=bash,caption={Cast/Migration Syntax}]
BIGDAWG_CAST ::=
  bdcast( BIGDAWG_RETRIEVAL_SYNTAX, name_of_intermediate_result, {
    {, POSTGRES_SCHEMA_DEFINITION, relational}
    | {, SCIDB_SCHEMA_DEFINITION, array}
    | {, TEXT_SCHEMA_DEFINITION, text}} )
\end{lstlisting}




The following is an example of moving data between array and
relational islands:

\begin{lstlisting}[language=bash,frame=none]
bdarray(
  scan(
    bdcast(
      bdrel(SELECT poe_id, subject_id FROM mimic2v26.poe_order LIMIT 5)
      , poe_order_copy
      , '<subject_id:int32>[poe_id=0:*,10000000,0]'
      , array)))
\end{lstlisting}

This query moves data from PostgreSQL to SciDB. The \texttt{bdrel()} portion of
the query selects the columns \texttt{poe\_id} and \texttt{subject\_id} from 
table \texttt{mimic2v26.poe\_order}. The \texttt{bdcast()} portion of the query tells the
middleware to migrate this data to an array called \texttt{poe\_order\_copy} with
schema \texttt{<subject\_id:int32>[poe\_id=0:*,10000000,0]} in the array
island. The final \texttt{bdarray()} portion of the query scans and returns to the user 
this resultant array in SciDB.



\section{Query Performance Analysis}
\label{results}

To characterize BigDAWG queries, we analyze system log data at the DEBUG level with millisecond
precision. This data is not meant to benchmark raw
performance, since the experiments were run on a single laptop running
all database engines in Docker containers. Instead, the goal is to compare
relative performance of subtasks within a query and with respect to query complexity.

\begin{figure}[t]
\centering
\includegraphics[width=3.6in]{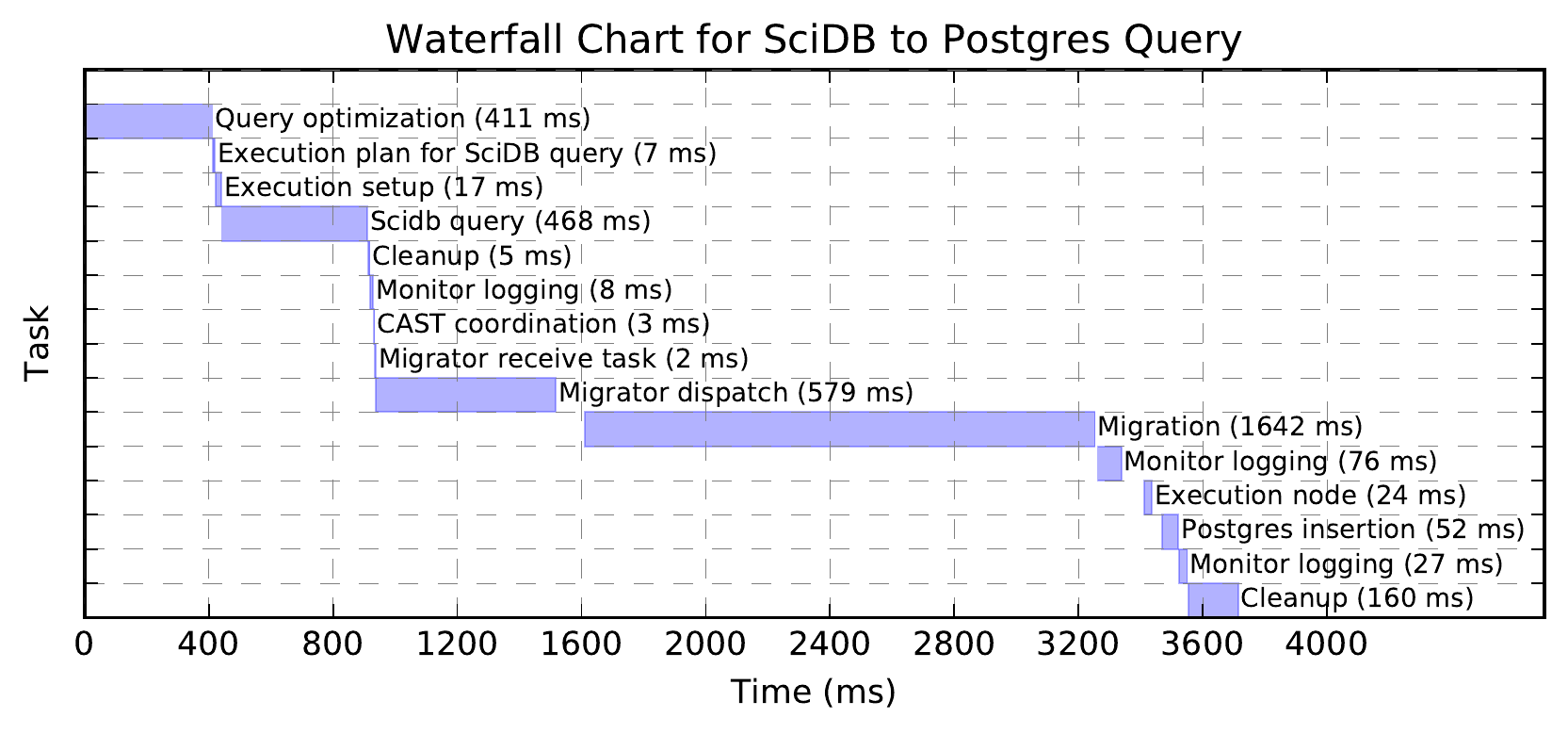}
\caption{Task timings for array and relational inter-island query.}
\label{waterfall}\end{figure}

Figure~\ref{waterfall} shows timings an inter-Island query between SciDB and Postgres. Tasks are
shown on the vertical axis and execution times (i.e. lengths of bars) on the horizontal axis. 

This plot shows the relative performance of each stage in
a complex query. Note that the bulk of the time is spent
waiting for two database execution tasks (\textit{Scidb query}), dispatching the remote
procedure call over the network(\textit{Migrator dispatch}), and transmitting data across the
network (\textit{Migration}). Together, these tasks consume
about 75\% of the total execution time. The overhead associated
with BigDAWG itself is mostly in the initial query optimization, which
is about 10\% of the total execution time. 

\begin{figure}[t]
\centering
\includegraphics[width=3in]{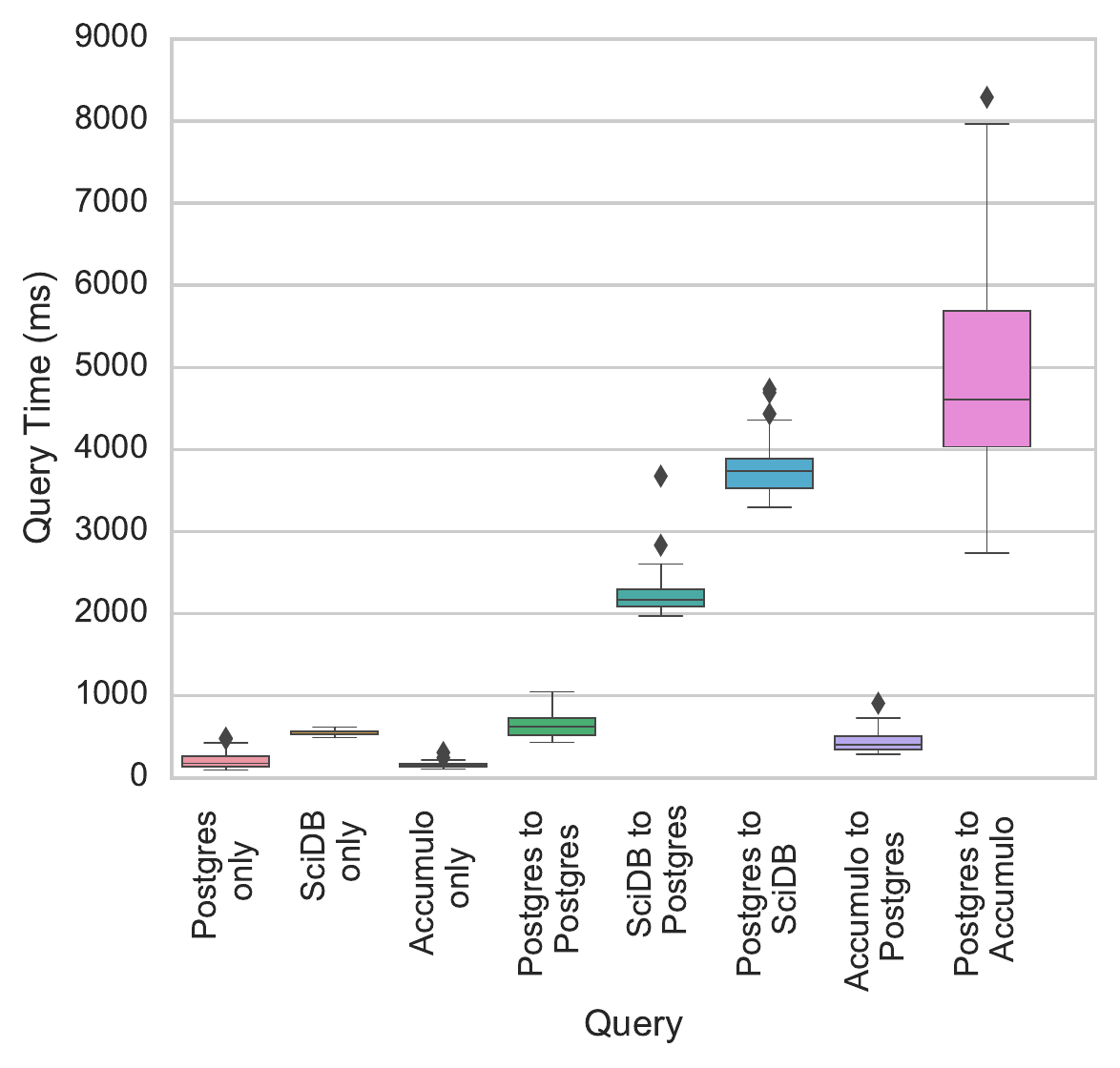}
\caption{Task timings for array and relational inter-island query.}
\label{boxplot}\end{figure}

Figure~\ref{boxplot} presents sample statistics for query execution times as a boxplot for eight
queries run 50 times.  The vertical axis shows the query execution time and
the horizontal axis shows the query type. This plot shows query latencies and speeds
for queries that require and queries that do not require migration between engines.  
As expected, queries that require migration take
more time than single island/engine queries. Inter-island migration between different island types 
require data format translation and transmission dispatching of remote procedures across the network, 
hence increasing query latency. However, migration between databases in the same island 
is fast since we can take advantage of native selection and insertion routines.





\section{Conclusion and Further Work}
\label{conc}

The future of data analytics will increasingly depend on data distributed 
among disparate database management
systems. Current practice based on federated database engines
cannot meet the needs of future high performance since they are largely limited to
single data or programming models. Polystore systems go 
well beyond data federation systems by supporting multiple query
languages and disparate yet integrated DBMSs. 

In this paper we described the open source release of our polystore system, BigDAWG version 0.1. 
As shown in prior results~\cite{gadepally2016bigdawg} with queries over the MIMIC II medical dataset,
BigDAWG provides dramatic performance advantages by using multiple storage engines 
optimized for particular operations and data models.  We build on that earlier study in this paper
showing that for complex queries, BigDAWG adds overhead of about 10\% of the total execution time. 

Future work on BigDAWG will expand its capabilities by adding additional islands
and storage engines. We plan to support more complex
query planning and additional execution capabilities.   We also plan
to improve inter-island query performance
by running monitor logging and cleanup tasks on background threads as well as using multithreaded 
execution for long-running tasks. Finally, we are interested in the
application of privacy preserving technologies, such as those presented
in~\cite{gadepally2015computing} and~\cite{fullersok}, to polystore databases.

\section*{Acknowledgement}
This work was supported in part by the Intel Science and Technology
Center (ISTC) for Big Data. The authors wish to thank our ISTC collaborators Kristin Tufte, Jeff Parkhurst, Stavros Papadopoulos,
Nesime Tatbul, Magdalena Balazinska, Bill Howe, Jeffrey Heer, David Maier,
Tim Kraska, Ugur Cetintemel, and Stan Zdonik. The authors also wish to
thank the MIT Lincoln Laboratory Supercomputing Center for their help in
maintaining the MIT testbed.

\begin{scriptsize}
\setlength{\parsep}{-1pt}\setlength{\itemsep}{0cm}\setlength{\topsep}{0cm}
\bibliographystyle{IEEEtran}
\bibliography{ref.bib}
\end{scriptsize}

\end{document}